\documentstyle[12pt]{article}
 \def\be{\begin{eqnarray}}
 \def\ee{\end{eqnarray}}
\input epsfig.sty
\begin{document} 
\pagestyle{empty}
\Huge{\noindent{Istituto\\Nazionale\\Fisica\\Nucleare}}

\vspace{-3.9cm}

\Large{\rightline{Sezione di ROMA}}
\normalsize{}
\rightline{Piazzale Aldo  Moro, 2}
\rightline{I-00185 Roma, Italy}

\vspace{0.65cm}

\rightline{INFN-1229/98}
\rightline{September 1998}

\vspace{1.cm}

\begin{center}{\large\bf Investigation of N-N*
Electromagnetic Form Factors within a Front-Form CQM}
\end{center}
\vskip 1em
\begin{center} E. Pace$^a$, G. Salm\`e$^b$ and S.  Simula$~^{c}$\end{center}

\noindent{$^a$\it Dipartimento di Fisica, Universit\`a di Roma
"Tor Vergata", and Istituto Nazionale di Fisica Nucleare, Sezione
Tor Vergata, Via della Ricerca Scientifica 1, I-00133, Rome,
Italy}

\noindent{$^b$\it Istituto Nazionale di Fisica Nucleare, Sezione
di  Roma I, P.le A. Moro 2, I-00185 Rome, Italy} 

\noindent{$^c$\it Istituto 
Nazionale di Fisica Nucleare, Sezione
di Roma III, Via della Vasca Navale 84, I-00146 Rome, Italy}

\vspace{2.cm}

\begin{abstract}
The helicity amplitudes for the transitions $N-S_{11}$ and $N-S_{31}$ are  
 presented. The amplitudes have been obtained within
 our front-form CQM model, based on hadron  eigenstates of a relativistic 
 mass operator  and CQ current with Dirac and Pauli form factors.

\end{abstract}

\vspace{4.5cm}
\hrule width5cm
\vspace{.2cm}
\noindent{\normalsize{Proceedings of XVI European Few-Body 
Conference on "Few-Body Problems in Physics", Autrans, June 1998. 
To appear in {\bf Few-Body Systems}, Suppl. }}

\newpage
\pagestyle{plain}

\sloppy

\vspace{2cm}

\noindent Hadron electromagnetic (em) form factors have been recently 
investigated
within the front-form constituent quark (CQ) model of  \cite{nuc95}  for 
space-like
values of the four-momentum transfer.  The main features of the model are:  
i)
the use of  hadron eigenfunctions of a relativistic mass operator, that 
includes
an effective $q-q$  interaction  and  reproduces the hadron spectra 
for a large set of  quantum numbers \cite{CI}; ii) the use of a
one-body em  current operator containing phenomenological Dirac and Pauli
 form
factors for CQ's, which are determined by the request of
reproducing the existing experimental data for the pion and nucleon elastic 
form
factors (cf. \cite{nuc95}).  Such a model has been already applied for 
obtaining  a parameter-free prediction of the em form factors for the 
  transitions to  $N^*(1440)$ and $\Delta(1232)$, including the possible 
  effects
due to the $D$-wave components in the $\Delta$ wave function, \cite{nuc95}.

In this contribution, we will present an analysis of transition form factors 
for
$N \rightarrow S_{11}(1535)$, $N \rightarrow S_{11}(1650)$ and $N \rightarrow 
S_{31}(1620)$.

The  current for negative-parity transition with $J_f=1/2$ is given in terms 
of
 Dirac ($F^{f\tau}_1$) and Pauli-like ($F^{f\tau}_2$) form factors by 
 (cf. \cite{Weber})
\begin{eqnarray}
\bar{\Psi}_f~J^{\mu}~\Psi_{\tau}=\bar{\Psi}_f\gamma^5 \left [ {p_f^{\mu}+
p_i^{\mu} \over M_f-M_i} F^{f\tau}_2 -{M_f+M_i \over M_f-M_i} q^{\mu}
F^{f\tau}_1 + \gamma^{\mu}  (F^{f\tau}_1+F^{f\tau}_2 ) \right ]\Psi_{\tau}
\label {cur}
\end{eqnarray}
where  $\tau =p,n$.
By using such a current, the helicities for negative-parity transition can be 
written as follows
\begin{eqnarray}
&S_{1/2}^{\tau}\left (Q^2 \right )= \zeta~\sqrt{{ 2 \pi \alpha \over k^*}} 
\sqrt{{ Q^+ \over 2 M_i M_f}}\sqrt{{ Q^+ Q^-\over 4  M_f}} {M_f- M_i \over Q^2 
\sqrt{2}} \left [F^{f\tau}_1- {Q^2 \over (M_f- M_i)^2} F^{f\tau}_2 \right ] 
\nonumber \\
&A_{1/2}^{\tau}\left (Q^2 \right )= -\zeta~\sqrt{{ 2 \pi \alpha \over k^*}} 
\sqrt{{ Q^+ \over 2 M_i M_f}}   \left (F^{f\tau}_1+ F^{f\tau}_2 \right )
\label{hel}
\end{eqnarray}
where $\zeta $ is the sign of the $\pi N$ decay amplitude, $k^*= 
(M^2_f -M^2_i)/2M_f$, $Q^{\pm}= (M_f\pm M_i)^2 +Q^2$. The invariant form 
factors in Eq. (\ref{hel}) can be obtained within the front-form  CQ model 
following standard procedures (see, e.g.,  \cite{nuc95}), namely approximating  
the plus component of the transition current, $\cal{I}^+$, in terms of the 
sum of one-body CQ currents, containing CQ Dirac and Pauli form factors. In 
particular
\begin{eqnarray}
F_1^{f\tau}= -{1 \over 2} Tr \left (\sigma_z {\cal I}^+(\tau) \right ) 
~~~~~~~~F_2^{f\tau}= -{M_f- M_i \over 2Q} Tr \left (\sigma_x {\cal I}^+(\tau) 
\right ).
\label{ff}
\end{eqnarray}
where ${\cal{I}}^{+}_{\nu_f \nu_i}(\tau)=\bar{u}^f_{LF}(\nu_f)\sum_{j=1}^3 ~ 
\left ( e_j \gamma^+ f_1^j(Q^2) ~ + ~ i
\kappa_j {\sigma^{+ \rho} q_{\rho} \over 2 m_j}f_2^j(Q^2) \right ) 
u^{\tau}_{LF}(
\nu_i)$.

In Figs. 1-5, our {\em parameter-free} evaluation of the helicity amplitudes, 
$A_{1/2}$ and $S_{1/2}$ are shown for $N \rightarrow S_{11}(1535)$, 
$S_{11}(1650)$ and $S_{31}(1620)$, respectively. In the case of $S_{31}(1620)$ 
the results for $p$ and $n$ coincides (as in the case of $P_{33}(1232)$), 
since only the isovector part of the CQ current is effective, given the 
isospin of the resonance.

\begin{figure}


\epsfig{bbllx=10mm,bblly=225mm,bburx=0mm,bbury=288mm,file=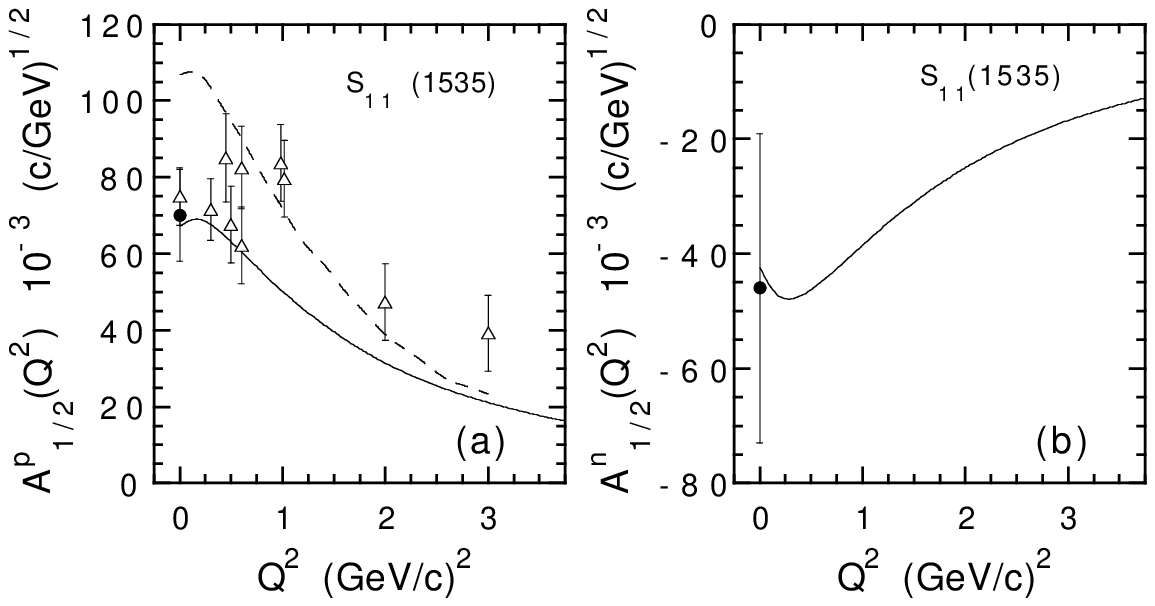}

{\bf Figure 1.} - (a) The transverse helicity $A_{1/2}$ for the transition 
$p \rightarrow S_{11}(1535)$ vs. $Q^2$. Solid line: $A_{1/2}$
from the hadron wave functions 
corresponding to the  interaction of  \cite{CI} and  the nucleon em 
current with  CQ form factors of \cite{nuc95};  dashed line:   a non 
relativistic CQM calculation \cite{Giannini}. Solid dot: PDG '96 \cite{PDG}; 
triangles: data analysis from \cite{Volker}. - (b) The same as in Fig. 1(a),
 but for
$n \rightarrow S_{11}(1535)$.

\end{figure}
\begin{figure}

\epsfig{bbllx=10mm,bblly=225mm,bburx=0mm,bbury=288mm,file=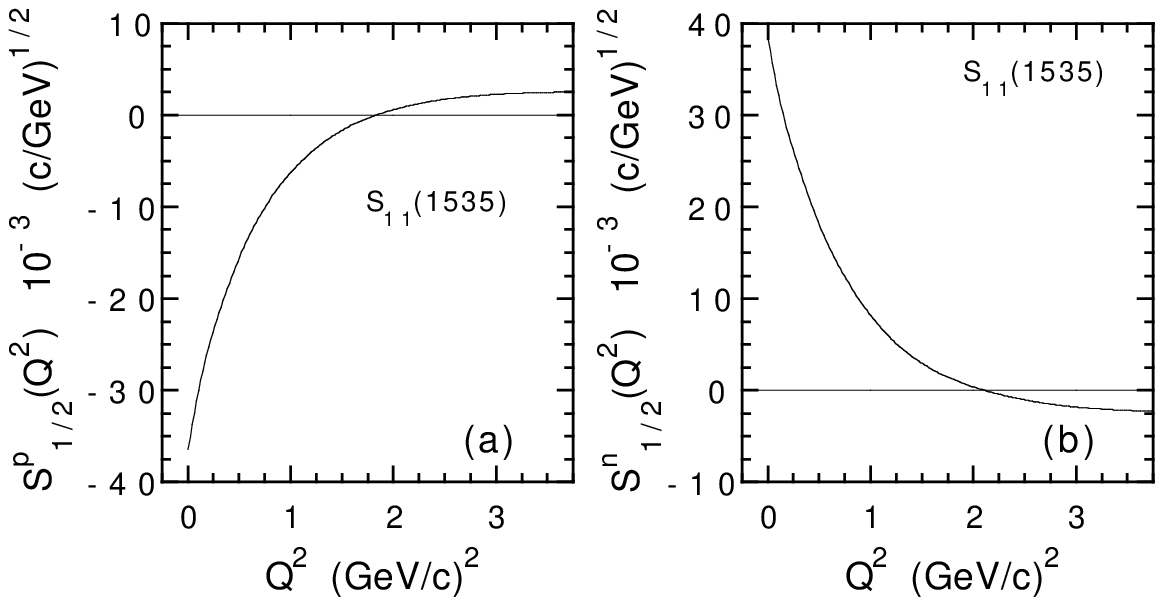}

{\bf Figure 2.} The same as in Fig. 1, but for the longitudinal helicity 
$S_{1/2}$. 

\end{figure}

\begin{figure}

\epsfig{bbllx=10mm,bblly=225mm,bburx=0mm,bbury=288mm,file=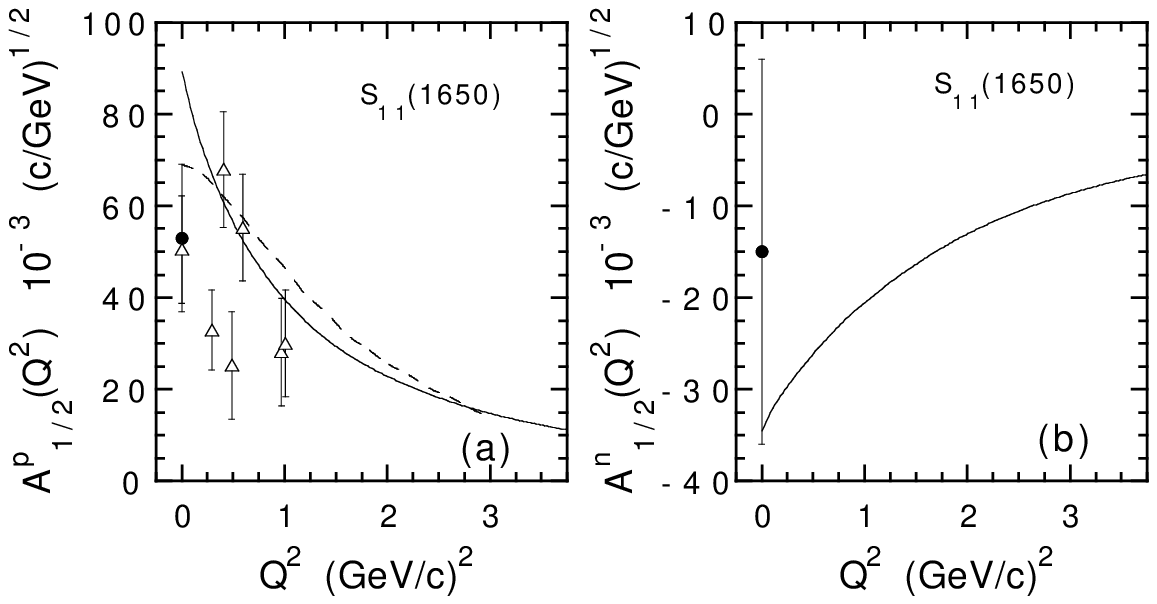}

{\bf Figure 3.} - (a) The transverse helicity $A_{1/2}$ for the transition 
$p  \rightarrow S_{11}(1650)$ vs. $Q^2$. Solid  line:  $A_{1/2}$
from the hadron wave functions 
corresponding to the  interaction of  \cite{CI} and  the nucleon em 
current with  CQ form factors of \cite{nuc95};  dashed line:   a non 
relativistic CQM calculation \cite{Giannini}. Solid dot: PDG '96 \cite{PDG}; 
triangles: data analysis from \cite{Volker}. - (b) The same as in Fig. 3(a), 
but 
for
$n \rightarrow S_{11}(1650)$.

\end{figure}

\begin{figure}

\epsfig{bbllx=10mm,bblly=225mm,bburx=0mm,bbury=288mm,file=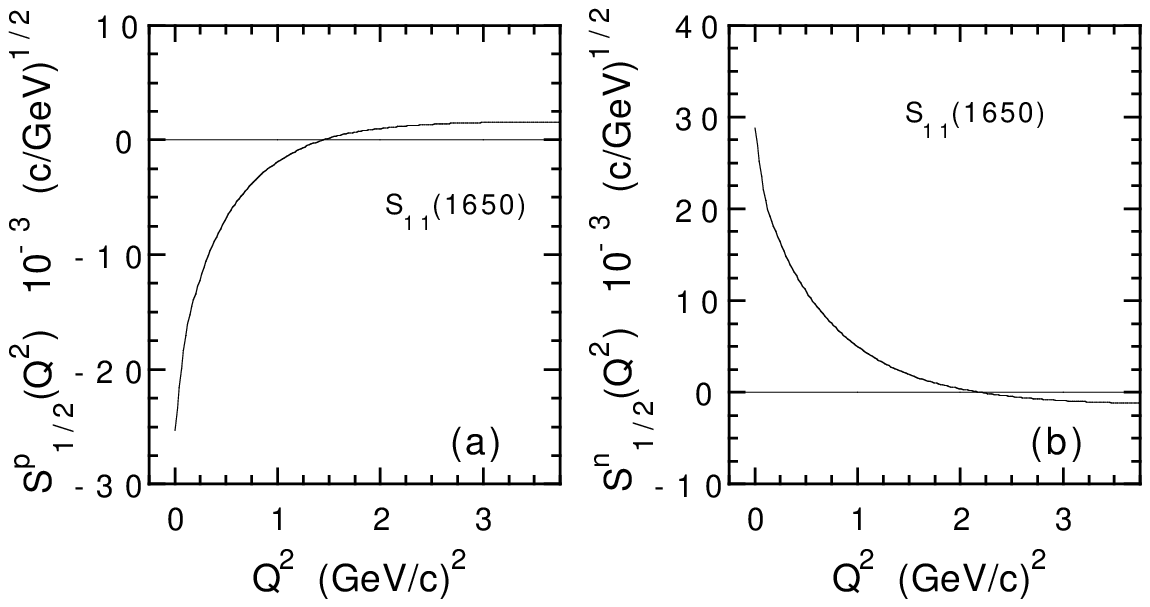}

{\bf Figure 4.} The same for Fig. 3, but for the longitudinal helicity 
$S_{1/2}$.

\end{figure}
\begin{figure}
\epsfig{bbllx=10mm,bblly=225mm,bburx=0mm,bbury=288mm,file=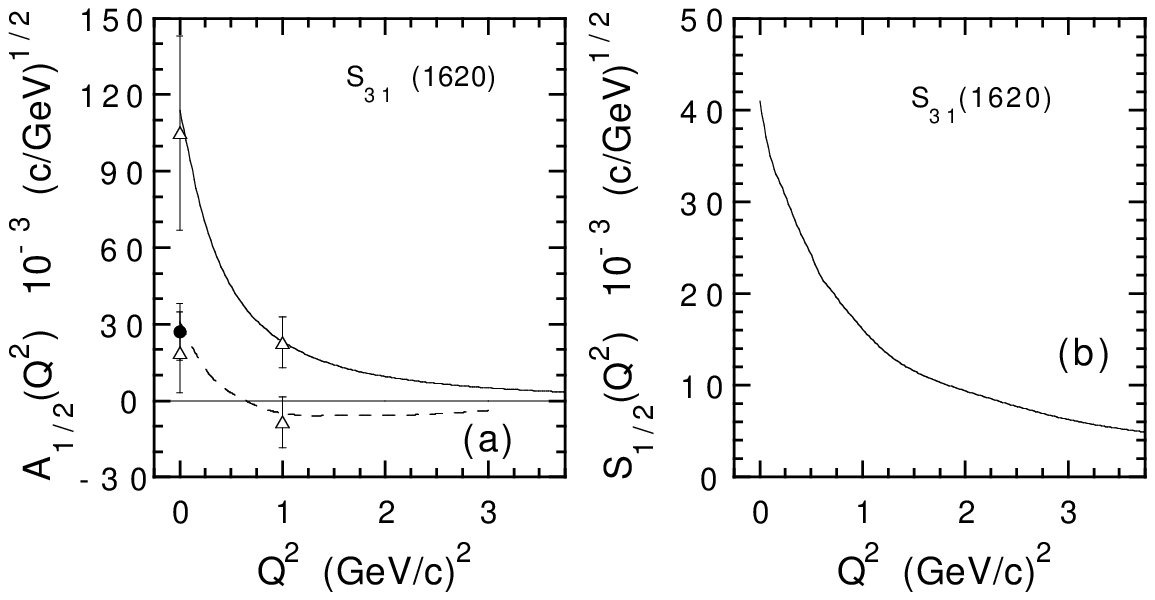}

{\bf Figure 5.} - (a) The transverse helicity $A_{1/2}$ for the transition 
$p \rightarrow S_{31}(1620)$ vs. $Q^2$. Solid line: $A_{1/2}$
from the hadron wave functions 
corresponding to the  interaction of  \cite{CI} and  the nucleon em 
current with  CQ form factors of \cite{nuc95}; dashed line:   a non 
relativistic CQM calculation \cite{Giannini}. - (b) The same as in Fig. 5a,
 but for $S_{1/2}$. 

\end{figure}

The overall agreement between our predictions and the data is encouraging, 
though a most accurate set of data is necessary in order to reliably 
discriminate  between  different models. However, the sensitivity to 
relativistic effects for the P-wave resonances seems sizable.

 \end{document}